\documentclass[twoside]{article}
\usepackage{amsmath}

\title{
\Large
{\bf On the integrability of nonlinear partial differential equations} 
}
\author{ {\bf H.J.S. Dorren} \\
{\small Department of Electrical Engineering, 
Eindhoven University of Technology,
P.O. Box 513,} \\
{\small 5600 MB Eindhoven, The Netherlands} }
\date{}

\begin{document}
\maketitle

\begin{abstract}
We investigate the integrability of 
Nonlinear Partial Differential Equations (NPDEs). 
The concepts are developed by firstly discussing the 
integrability of the KdV equation. We proceed by generalizing the ideas 
introduced for the KdV equation to other NPDEs. 
The method is based upon a linearization principle which can be
applied on nonlinearities which have a polynomial form.
We illustrate the potential of the method  by finding solutions of the
(coupled) nonlinear Schr\"{o}dinger equation and the Manakov equation 
which play an important role in optical fiber communication. 
Finally, it is shown that the method can also  be generalized to 
higher-dimensions. 
\end{abstract}

\vspace{12cm}
\begin{flushleft}
\begin{small}
\hspace{1cm}
E-mail: H.J.S.Dorren@ele.tue.nl 
\end{small}
\end{flushleft}

\newpage
\section{Introduction}
The conditions under which Nonlinear Partial Differential Equations (NPDEs)
can be solved are even in one dimension not well understood \cite{Zakharov}. 
Roughly speaking the majority of the integrable systems can be classified in 
three main groups. In the first of these groups are those equations which can 
be reduced to a quadrature through the existence of an adequate number 
of integrals of motion. In the second class are those equations which can 
be mapped into a linear system by applying a number of transformations
(hereafter to be called C-integrable) \cite{Eck1}. The last group 
consists of differential equations which can be solved by 
Inverse Scattering Transformations (IST). In the following, we will call 
equations which can be solved by inverse scattering methods
``S-integrable''. The discovery of the IST has lead to a considerable progress 
in understanding the topic of integrability since this 
technique made it possible to investigate the integrability of large 
classes of NPDEs systematically \cite{Ablowitz}. Recently, it has been 
discovered that NPDEs which can be solved by inverse scattering techniques 
can also be transformed into linear differential equations \cite{Dorren}. 
We therefore have the feeling that NPDEs which can be solved by inverse 
scattering methods form a sub-class of the C-integrable NPDEs.

Another important consideration is that almost all the work on the 
integrability of NPDEs has been carried out in one space dimension only. 
Although the inverse problem of the Schr\"{o}dinger equation can be 
generalized to three dimensions, the method is far too complicated to solve
higher dimensional NPDEs. An alternative is the
$\overline{\partial}$-approach which is also successfully generalized to 
N-dimensions (see for instance Ref.\cite{Ablowitz}). Nevertheless, for both 
these methods the existence of the obtained solutions is difficult to prove. 
The concept of C-integrability however, has the potential to be 
generalized easily to dimensions higher than one. In this paper, we will 
demonstrate a simple method 
based upon linearization principles that generalizes the concepts of 
C-integrability and S-integrability and enables us to compute 
general solutions of large classes of NPDEs by solving a linear algebraic 
recursion relationship. The result suggests that this method can also be 
generalized to systematically derive large classes of solutions
of higher-dimensional NPDEs.      

In this paper we aim to find large classes of integrable equations which 
can be solved by linearization.  Since it is not clear what integrability 
exactly means, we use in this paper the heuristic definition that an NPDE 
is integrable if given a sufficiently general initial condition, we can find
analytic expressions the time-evolution of the solution. For NPDEs which can
be solved by inverse scattering techniques, this notion is equivalent with the 
existence of N-soliton solutions, since it is implicitly assumed that the 
obtained solution can be expanded on a Fourier basis \cite{Dorren}.
In this paper, we will show that the condition of expansion in a Fourier can 
be replaced by an arbitrary other complete set of basis-functions. 

We present the following novel results. Firstly, we derive a simple method to
find general solutions of large classes of NPDEs. Secondly, we show that 
the integrability of these NPDEs is guaranteed if the nonlinearity 
can be expanded in the same basis-functions as the linear part, 
and secondly, if the dispersion relationship associated with the linearized 
problem can be solved. The method is firstly illustrated by discussing the 
integrability of the KdV equation in Section 2. 
In Section 3, the concepts derived for the KdV equation are 
generalized to discuss the integrability of general NPDEs. Finally, 
in Section 4, the results are applied to investigate the integrability of the 
coupled nonlinear Schr\"{o}dinger equation and the Manakov problem which play 
an important role in the field of optical fiber communications. 
The paper is concluded with a discussion.

\section{The integrability of the KdV equation}
In order to illustrate the machinery developed throughout this paper we 
firstly discus 
the integrability of the KdV equation as an example. The integrability of the 
KdV equation is a well-studied problem \cite{Ablowitz}. This makes the KdV equation an ideal
object to test the validity of newly developed ideas with respect to 
the integrability of NPDEs. We will introduce our methods on the integrability
of NPDEs by discussing the existence of N-soliton solutions for the 
KdV equation which is given by:
\begin{equation}
u_{t} +  u_{xxx} = 6 u_{x} u
\label{kdv}
\end{equation}
We try to find solutions of Eq.(\ref{kdv}) by substitution of  
the following Fourier series:
\begin{equation}
u(x,t)= \sum_{n=1}^{\infty} A_{n} e^{ in (kx-\omega t) }
\label{series}
\end{equation} 
If we substitute the solution $u(x,t)$ into Eq.(\ref{series}), we obtain:
\begin{equation}
\sum_{n=1}^{\infty} 
\left( 
n \omega + k^{3} n^{3} 
\right)
A_{n} e^{ in (kx-\omega t) } 
= -6 k
\sum_{n=1}^{\infty} \sum_{l=1}^{n-1}  
l  A_{l} A_{n-l} e^{ in(kx-\omega t) }
\label{ser_sub}
\end{equation}
We can now determine the coefficients $A_{n}$ by deriving a recursion 
relationship. This can be achieved by comparing the exponential 
functions in Eq.(\ref{ser_sub}). If we compare all the terms for which 
$n=1$, we find:
\begin{equation}
\left(
\omega + k^{3}
\right) A_{1} e^{ i(kx-\omega t) } = 0.
\label{kdvn=1}
\end{equation}
For a nonzero $A_{1}$, we find that Eq.(\ref{kdvn=1}) is satisfied if:
\begin{equation}
\omega = - k^{3}
\label{n=1term}
\end{equation}
If we put $n=2$, we can determine $A_{2}$ by solving the following 
relationship:
\begin{equation}
\left(
2 \omega + 8  k^{3}
\right) 
A_{2} e^{ 2 i (kx-\omega t) } 
= 
- 6 k A_{1} A_{1} 
e^{ 2 i(kx-\omega t) }
\end{equation}
If we use the dispersion relationship (\ref{n=1term}), we find 
that $A_{2}$ is given by: 
\begin{equation}
A_{2}  
= - \frac{
A^{2}_{1}
}{
k^{2}
}
\end{equation}
By repeating this procedure, we can compute all the expansion coefficients
$A_{n}$ 
of the solutions $u(x,t)$. In general, all the coefficients $A_{n}$ can 
be computed by solving the following linear algebraic problem:
\begin{equation}
L^{(n)}(k) A_{n} = R^{(n)}(k)
\label{alg_rel}
\end{equation}
The operators $L^{(n)}(k)$ and $R^{n}(k)$ in Eq.(\ref{alg_rel}) are
given by:
\begin{equation}
L^{(n)}(k) = n [n^{2} - 1]k^{3}
\hspace{1cm}
R^{(n)}(k) = - 6 k \sum_{l=1}^{n-1}  l  A_{l} A_{n-l}   
\label{def_rel}
\end{equation}
If we compute all the coefficients $A_{n}$ by using Eq.(\ref{def_rel}), we 
than obtain the Fourier expansion of $u(x,t)$ for which the first terms are 
given by:
\begin{equation}
u(x,t) = A_{1} e^{i(k x - \omega t) } - 
\frac{A^{2}_{1}}{k^{2}}
e^{2i(k x - \omega t) } + 
\frac{3 A^{3}_{1}}{ 4 k^{4}} e^{-3i( k x -  \omega t) } 
+ \cdots
\label{sol_series}
\end{equation}
If we substitute $k= 2 i \beta$ and $A_{1} = 4 d \beta$ into
Eq.(\ref{sol_series}) we find:
\begin{equation}
u(x,t) = 4d \beta e^{-2(\beta x - 4 \beta^{3} t) } +          
16 d^{2} e^{-4(\beta x - 4 \beta^{3} t) } +
\frac{24 d^{3} }{ \beta } 
e^{-6(\beta x - 4 \beta^{3} t) } 
+ \cdots
\label{ser_soliton}
\end{equation}
By carrying out the summation in Eq.(\ref{ser_soliton}), we can formulate
this equation more compactly:
\begin{equation}
u(x,t) =
\frac{
8 d \beta e^{-2(\beta x - 4 \beta^{3} t) }
}{
\left( 
1 + \frac{d}{\beta} e^{-2(\beta x - 4 \beta^{3} t) }
\right)^{2}
}
\label{ser_soliton1}
\end{equation}
Hence, if we put:
\begin{equation}
\beta= \frac{1}{2} \sqrt{c},
\hspace{1cm}
x_{0} = -  \frac{1}{\sqrt{c}} \log \left( - \frac{d}{\beta} \right),
\hspace{1cm}
d < 0,
\end{equation}
we can simplify Eq.(\ref{ser_soliton1}) one step further to:
\begin{equation}
u(x,t) = 
- 
\frac{c}{2} \mbox{sech}^{2} \left\{
\frac{1}{2} \sqrt{c} ( x - ct + x_{0} ) \right\}.
\label{kdv_sol}
\end{equation}
Eq.(\ref{kdv_sol}) describes the well-known KdV soliton.

What did we learn from this simple exercise? At first, the KdV equation
has solutions because of the special structure of the nonlinearity. If we 
substitute the special solution (\ref{series}) in the nonlinear part of the
KdV equation, we find that we can expand the nonlinearity in the same basis
functions as the linear part:
\begin{equation}
6 u_{x} u = \sum_{n=1}^{\infty} D_{n} e^{in(kx-\omega t)};
\hspace{1cm}
D_{n} = - 6 k \sum_{l=1}^{n-1}  l A_{l} A_{n-l}
\label{nlin}
\end{equation}
This guarantees that we can find an iteration relationship for the expansion 
coefficients $A_{n}$. As we will see later, we do not have to restrict to a 
Fourier expansion of the solution only. In principle this method works for any 
set of basis-functions as long as we can expand the nonlinearity in the 
same basis-functions as the linear part. In the following, we will show that 
the structure of the nonlinearity of the KdV equation enables us to construct
the Fourier expansion of the N-soliton of the KdV equation. In order to  
systematically solve these solutions it is illustrative to discus 
also the two-soliton solutions which are assumed to have the following series 
expansion:
\begin{equation}
u(x,t) =
\sum_{\mu_{1},\mu_{2}=1}^{\infty} C(\mu_{1},\mu_{2})
e^{i (\mu_{1} k_{1} z_{1}+ \mu_{2} k_{2} z_{2} ) }
\hspace{1cm}
\left\{
\begin{array}{c}
z_{1} = x - \frac{ \omega(k_{1}) }{ k_{1} } \\
z_{2} = x - \frac{ \omega(k_{2}) }{ k_{2} }
\end{array}
\right.
\label{try_2sol}
\end{equation}
If we substitute Eq.(\ref{try_2sol}) into the KdV equation (\ref{kdv}),
we obtain the following result:
\begin{equation}
\begin{split}
&\sum_{\mu_{1},\mu_{2}=1}^{\infty} 
L^{(\mu_{1},\mu_{2})}(k_{1},k_{2}) C(\mu_{1},\mu_{2})
e^{i (\mu_{1} k_{1} z_{1}+ \mu_{2} k_{2} z_{2} ) } \\ 
&= -6
\sum_{\mu_{1},\mu_{2}=1}^{\infty}
\sum_{\eta_{1},\eta_{2}=1}^{\mu_1-1,\mu_{2}-1}
M^{(\eta_{1},\eta_{2})}(k_{1},k_{2}) 
C(\mu_{1}-\eta_{1},\mu_{2}-\eta_{2}) C(\eta_{1},\eta_{2})   
e^{i (\mu_{1} k_{1} z_{1}+ 
      \mu_{2} k_{2} z_{2} ) }
\label{ser_sub2}
\end{split}
\end{equation}
where
\begin{equation}
L^{(n_{1},n_{2})}(k_{1},k_{2}) = 
\sum_{i=1}^{2} n_{i}[n_{i}^{2}-1]k_{i}^{3}
\hspace{1cm}
M^{(n_{1},n_{2})}(k_{1},k_{2}) = 
\sum_{i=1}^{2} n_{i} k_{i}
\end{equation}
We solve Eq.(\ref{ser_sub2}) by comparing equal exponential powers on both 
sides. This can be done by defining a parameter $\Gamma=\mu_{1}+\mu_{2}$ 
and subsequently comparing the powers for $\Gamma=1,2,3, \cdots$. 
We firstly discus the case in which $\Gamma=1$ in which only the 
coefficients $C(1,0)$ and $C(0,1)$ contribute:
\begin{equation}
\left[ \omega_{1} + k_{1}^{3} \right]C(1,0) e^{ik_{1}z_{1}} +
\left[ \omega_{2} + k_{2}^{3} \right]C(0,1) e^{ik_{2}z_{2}} = 0
\end{equation} 
If we put $C(1,0)=A_{1}$ and $C(0,1)=A_{2}$, we find that the 
following linear dispersion relationships must be valid: 
\begin{equation}
\omega(k_{1}) = - k^{3}_{1}
\hspace{1cm}
\mbox{and}
\hspace{1cm}
\omega(k_{2}) = - k^{3}_{2}
\label{lin_def}
\end{equation}
Once the linear dispersion relationships are determined and if the
coefficients $C(1,0)$ and $C(0,1)$ have taken their values $A_{1}$ 
and $A_{2}$, we can compute all the other coefficients $C(\mu,\eta)$ 
by applying the following linear recursion relation:
\begin{equation}
L^{(\mu_{1},\mu_{2})}(k_{1},k_{2}) 
C(\mu_{1},\mu_{2}) = R^{(\mu_{1},\mu_{2})}(k_{1},k_{2})
\label{it_2} 
\end{equation}
where
\begin{equation}
R^{(\mu_{1},\mu_{2})}(k_{1},k_{2}) =
- 6 \sum_{\eta_{1},\eta_{2}=1}^{\mu_1-1,\mu_{2}-1}
M^{(\eta_{1},\eta_{2})}(k_{1},k_{2}) 
C(\mu_{1}-\eta_{1},\mu_{2}-\eta_{2}) C(\eta_{1},\eta_{2})   
\end{equation}
Equation (\ref{it_2}) has a similar structure as Eq.(\ref{alg_rel}).  
In principle Eq.(\ref{it_2}) provides an efficient tool to compute 
all the coefficients $C(\mu,\eta)$. We can easily generalize this
result to the N-soliton case by assuming that the solution $u(x,t)$ takes 
the following form:
\begin{equation}
u(x,t) =
\sum_{\mu_{1} \cdots \mu_{N}=1}^{\infty} C(\mu_{1}\cdots \mu_{N})
e^{i (\mu_{1} k_{1} z_{1}+ \cdots +\mu_{N} k_{N} z_{N} ) }
\hspace{1cm}
\left\{
\begin{array}{c}
z_{1} = x - \frac{ \omega(k_{1}) }{ k_{1} } \\
\vdots \\
z_{N} = x - \frac{ \omega(k_{N}) }{ k_{N} }
\end{array}
\right.
\label{try_nsol}
\end{equation}
We can determine the non-zero coefficients $C(\mu_{1}\cdots \mu_{N})$ by 
substituting Eq.(\ref{try_nsol}) into the KdV equation (\ref{kdv}):
\begin{equation}
\begin{split}
&\sum_{\mu_{1} \cdots \mu_{N}=1}^{\infty}
\! \! \! \! \! \! 
L^{(\mu_{1} \cdots \mu_{N})}(k_{1} \cdots k_{N}) 
C(\mu_{1} \cdots \mu_{N})
e^{i (\mu_{1} k_{1} z_{1}+ \cdots + \mu_{N} k_{N} z_{N} ) } \\ 
&= -6 
\sum_{\mu_{1} \cdots \mu_{N}=1}^{\infty}
\sum_{\eta_{1} \cdots \eta_{N}=1}^{\mu_{1}-1 \cdots \mu_{N}-1 }
M^{(\eta_{1} \cdots \eta_{N})}(k_{1} \cdots k_{N}) 
C(\mu_{1}-\eta_{1} \cdots \mu_{N} - \eta_{N}) 
C(\eta_{1} \cdots \eta_{N})   
e^{i (\mu_{1} k_{1} z_{1}+ \cdots + \mu_{N} k_{N} z_{N} ) }
\end{split}
\label{ser_subn}
\end{equation}
where
\begin{equation}
L^{(n_{1} \cdots n_{N})}(k_{1} \cdots k_{N}) 
= \sum_{i=1}^{N} n_{i}[n_{i}^{2}-1]k_{i}^{3};
\hspace{1cm}
M^{(n_{1} \cdots n_{N})}(k_{1} \cdots k_{N})
= \sum_{i=1}^{N} n_{i}k_{i} 
\end{equation}
If we use that $\omega(k_{i}) = - k^{3}_{i}, ( i \in 1 \cdots N)$ and  
$A_{1}=C(1,0,0,\cdots,0), A_{2}=C(0,1,0,\cdots,0), 
\cdots ,A_{N}=C(0,\cdots,0,1)$, 
we find that the expansion coefficients of the N-soliton solution for the 
KdV equation can be computed by solving the following linear relationship:
\begin{equation}
L^{(\mu_{1} \cdots \mu_{N})} (k_{1} \cdots k_{N}) 
C(\mu_{1} \cdots \mu_{N} ) = R^{(\mu_{1} \cdots \mu_{N})}(k_{1} \cdots k_{N})
\label{lin_shape}
\end{equation}
where
\begin{equation}
R^{(\mu_{1} \cdots \mu_{N})}(k_{1} \cdots k_{N}) = - 6
\sum_{\eta_{1} \cdots \eta_{N}=1}^{\mu_{1}-1 \cdots \mu_{N}-1 }
M^{(\eta_{1} \cdots \eta_{N})}(k_{1} \cdots k_{N}) 
C(\mu_{1}-\eta_{1} \cdots \mu_{N} - \eta_{N}) 
C(\eta_{1} \cdots \eta_{N})   
\label{linform} 
\end{equation} 
From the exercise performed in this section we can conclude that general
solutions of the KdV equation can be obtained by solving 
Eqs.(\ref{lin_shape}). This implies that the KdV equation 
can be transformed into a simple linear algebraic equation in the coefficient 
space. We can conclude that the KdV equation has N-soliton solutions 
because the following two conditions are satisfied:  
\begin{itemize}
\item{The structure of the nonlinearity of the
KdV equation guarantees that the equation has solutions of the form 
(\ref{try_nsol}). This result implies that the coefficients
$R^{(\mu_{1} \cdots \mu_{N})}(k_{1} \cdots k_{N})$ exist.}
\item{$L^{(n_{1} \cdots n_{N})}(k_{1} \cdots k_{N})$
is not equal to zero if $k_{1} \cdots k_{N} \neq 0$ and 
$n_{1} \cdots n_{N} \neq 0$. This implies that 
$L^{(n_{1} \cdots n_{N})}(k_{1} \cdots k_{N})$ has an inverse.}
\end{itemize} 
In the following section we will show that a similar condition must 
hold for more general nonlinear NPDEs. In the following section it is 
shown that the concepts derived for the KdV equation can be generalized 
to large classes of NPDEs. The results obtained in this section 
are derived by assuming that the solution of the KdV equation 
can be expanded in Fourier basis functions. In the following section, it will
be shown that similar principles apply for general basis functions.

\section{Generalizations}
In this section we will present more general results with respect to the 
integrability of nonlinear evolution equations. This will be done by 
generalizing the results obtained for the KdV equation.  
In this section, we focus on NPDEs of the following type:
\begin{equation}
{\cal L}[{\bf u}(x,t)] = Q[{\bf u}(x,t)]
\label{prob}
\end{equation}
In Eq(\ref{prob}), the function ${\bf u}(x,t)$ is a M-component vector 
function having entries $u_{i}(x,t)$. The operator ${\cal L}[ \ \cdot \ ]$ is assumed
to
take the following form:
\begin{equation}
{\cal L}[{\bf u}(x,t)] = 
\left[ i {\bf I} \frac{\partial}{\partial t} + 
\sum_{n=1}^{K} {\bf A}^{(n)} \frac{\partial^{n}}{\partial x^{n}}
\right]  {\bf u}(x,t)
\label{A_rep}
\end{equation}
The matrices ${\bf A}^{(n)}$ in Eq.(\ref{A_rep}) are $M \times M$ matrices
and ${\bf I}$ is the identity matrix.
As concluded from the previous section, integrability puts strong constraints
on the nonlinearity represented by the operator $Q$. As a necessary condition 
for the integrability we require that if a solution of Eq.(\ref{prob}) has 
the following form:
\begin{equation}
{\bf u}(x,t) =
\sum_{ \mu_{1} \cdots \mu_{N}=1}^{\infty} 
{\bf C}(\mu_{1} \cdots \mu_{N})
\exp 
\left[
i \sum_{r=1}^{N} \sum_{s=1}^{M}  \mu_{r} k_{rs} z_{rs} 
\right];
\hspace{1cm}
z_{rs} = x - \frac{
\omega(k_{rs})
}{
t
},
\label{test1}
\end{equation}
than, the operator $Q$ must satisfy the following property: 
\begin{equation}
Q[{\bf u}(x,t)] =
\sum_{ \mu_{1} \cdots \mu_{N}=1}^{\infty} 
{\bf R}(\mu_{1} \cdots \mu_{N})
\exp 
\left[
i \sum_{r=1}^{N} \sum_{s=1}^{M}  \mu_{r} k_{rs} z_{rs} 
\right],
\label{Q_opp}
\end{equation}
where ${\bf C}(\mu_{1} \cdots \mu_{N})$ and ${\bf R}(\mu_{1} \cdots \mu_{N})$
are M-dimensional vector functions. Similarly as for the KdV equation,    
the vector function ${\bf R}(\mu_{1} \cdots \mu_{N})$ is specified by the 
nonlinearity. In other words: we require that given a solution 
of the form (\ref{test1}), the nonlinear operator $Q[{\bf u}(x,t)]$ can be 
expanded in the same set of basis functions as ${\cal L}[{\bf u}(x,t)]$. In 
the previous section, we have shown that the nonlinearity of the KdV equation 
satisfies this condition. In general large classes of nonlinear operators 
will have the property (\ref{Q_opp}) and among them we are especially 
interested in the sub-class $\hat{P}$ which plays an important role in 
nonlinear optics:
\begin{equation}
\hat{P}[{\bf u}(x,t)] =
P_{N} \left( 
{\bf u}, \frac{\partial {\bf u}}{\partial x},
\frac{\partial {\bf u}}{\partial t},
\cdots
\frac{\partial^{p} {\bf u}}{\partial x^{q} \partial t^{p-q} }
\right),
\end{equation}
where $P_{N}$ are polynomials of order $N$.   
If we let act the the linear operator ${\cal L}$ onto the solution
(\ref{test1}) we obtain the following relationship:
\begin{equation}
{\cal L}[{\bf u}(x,t)] = \left( 
{\bf I} \sum_{r=1}^{N} \sum_{s=1}^{M}  \mu_{r} \omega(k_{rs}) +
\sum_{n=1}^{K}{\bf A}^{(n)} \left[ i 
\sum_{r=1}^{N} \sum_{s=1}^{M} \mu_{r} k_{rs} 
\right]^{n} 
\right)
{\bf u}(x,t)
\label{A_rep1}
\end{equation}
From this result, we can identify a matrix 
${\bf L}^{(\mu_{1} \cdots \mu_{N})}(k_{ij})$ which is given by:
\begin{equation}
{\bf L}^{(\mu_{1} \cdots \mu_{N})}(k_{ij}) =
{\bf I} \sum_{r=1}^{N} \sum_{s=1}^{M}  \mu_{r} \omega(k_{rs}) +
\sum_{n=1}^{K}{\bf A}^{(n)} \left[ i 
\sum_{r=1}^{N} \sum_{s=1}^{M} \mu_{r} k_{rs} 
\right]^{n}
\end{equation}
This result implies that the coefficients ${\bf C}(\mu_{1} \cdots \mu_{N})$ 
which determine the solution (\ref{test1}) can be determined by solving: 
\begin{equation}
{\bf L}^{(\mu_{1} \cdots \mu_{N})}(k_{ij}) 
{\bf C}(\mu_{1} \cdots \mu_{N})= {\bf R}(\mu_{1} \cdots \mu_{N})
\end{equation}
The coefficients ${\bf C}(1,0,0,\cdots,0), {\bf C}(0,1,0,\cdots,0), 
\cdots,{\bf C}(0,\cdots,0,1)$ are determined by the initial condition. 
 
In principle we expand the solution $u(x,t)$ in an arbitrary set of 
basis-functions. Suppose as an example a function $\hat{u}(x,t)$ which can 
be expanded in the set of basis-functions $f^{(n)}(x,t|k,\omega)$:
\begin{equation}
\hat{u}(x,t) = \sum_{n=1}^{\infty} \alpha_{n} f^{(n)}(x,t|k,\omega) 
\end{equation}
We define the set ${\cal S}$ as the basis-functions:
\begin{equation}
{\cal S} = \{ 
f^{(1)}(x,t|k,\omega),f^{(2)}(x,t|k,\omega),f^{(3)}(x,t|k,\omega), \cdots
\},
\end{equation}
which have the following properties:
\begin{equation}
\begin{split}
I&: \ \ \
\mbox{if} \ f^{(n)}(x,t|k,\omega) \ \in \ {\cal S} \Rightarrow 
\frac{\partial }{\partial t} f^{(n)}(x,t|k,\omega) = 
\hat{\alpha}_{n}(k,\omega) f^{(m)}(x,t|k,\omega) 
\hspace{1cm} (f^{(m)}(x,t) \ \in \ {\cal S}) \\
II&: \ \ \
\mbox{if} \ f^{(n)}(x,t|k,\omega) \ \in \ {\cal S} \Rightarrow 
\frac{\partial }{\partial x} f^{(n)}(x,t|k,\omega) = 
\hat{\beta}_{n}(k,\omega) f^{(m)}(x,t|k,\omega)
\hspace{1cm} (f^{(m)}(x,t) \ \in \ {\cal S}) \\
III&: \ \ \
\mbox{if} \ f^{(n)}(x,t) \ \in \ {\cal S}
\ \mbox{and} \
\ f^{(m)}(x,t) \ \in \ {\cal S} \ \Rightarrow \ 
f^{(n)}(x,t) \cdot f^{(m)}(x,t) \ \in \ {\cal S}.
\end{split}
\label{conditions}
\end{equation}
The properties $I$ and $II$ guarantee that 
${\cal L}[\hat{u}(x,t)]$ can be expanded in basis functions 
$f^{(n)}(x,t|k,\omega)$:
\begin{equation}
{\cal L}[\hat{u}(x,t)] = \sum_{n=1}^{\infty}
\hat{L}^{(n)} \alpha_{n} f^{(n)}(x,t|k,\omega)
\end{equation}
Where the precise structure of the operator $\hat{L}^{(n)}$ is determined
by the linear differential operator ${\cal L}$. Property $III$ in 
Eq.(\ref{conditions}) guarantees nonlinearities of the type $\hat{P}$ 
can be expanded in the same basis functions $f^{(n)}(x,t|k,\omega)$. If the 
nonlinearity represented by the operator $\hat{P}$ can also be expanded 
in the same basis-functions $f^{(n)}(x,t|k,\omega)$:
\begin{equation}
Q[\hat{u}(x,t)] = \sum_{n=1}^{\infty} \hat{R}_{n} f^{(n)}(x,t) 
\end{equation}
than, we can compute the expansion coefficients $\alpha_{n}$ by solving
the relationship:
\begin{equation}
\alpha_{n} = \left[ \hat{L}^{(n)}\right]^{-1} \hat{R}_{n}
\end{equation}
where $\alpha_{1}$ is determined by the initial condition. Of course, we can 
generalize this result further by replacing Eq.(\ref{test1}) by: 
\begin{equation}
{\bf u}(x,t) =
\sum_{ \mu_{1} \cdots \mu_{N}=1}^{\infty} 
{\bf \hat{C}}(\mu_{1} \cdots \mu_{N})
\prod_{i=1}^{N} \prod_{j=1}^{M}
f^{(i)}(x,t|\hat{k}_{ij},\hat{\omega}_{ij}) 
\label{test2}
\end{equation}
The structure of the solutions proposed in Eq.(\ref{test2}) is in fact a
generalization of Eq.(\ref{test1}). If we replace 
$f^{(i)}(x,t|\hat{k}_{ij},\hat{\omega}_{ij})$ by 
$\mbox{exp} \left[ i \mu_{i} k_{ij} z_{ij} \right]$, the form (\ref{test1}) 
is retained.
Following a similar approach as in the case of Fourier basis functions,
we find that if the conditions (\ref{conditions}) hold for the solution 
(\ref{test2}), the linear part of the differential equation acts on 
the solution (\ref{test2}) like:
\begin{equation}
{\cal L}[{\bf u}(x,t)] = \left( 
i {\bf I} \sum_{i=1}^{N} \sum_{j=1}^{M} \hat{\omega}_{ij} +
\sum_{n=1}^{K} {\bf A}^{(n)} 
\left[ \sum_{i=1}^{N} \sum_{j=1}^{M} \hat{k}_{ij} 
\right]^{n} 
\right)
{\bf u}(x,t)
\label{A_rep2}
\end{equation}
where it is assumed that 
$\partial_{t} f^{(i)}(x,t|\hat{k}_{ij},\hat{\omega}_{ij}) =
\hat{\omega}_{ij} f^{(i)}(x,t|\hat{k}_{ij},\hat{\omega}_{ij})$ 
and 
$\partial_{x} f^{(i)}(x,t) =
\hat{k}_{ij} f^{(i)}(x,t|\hat{k}_{ij},\hat{\omega}_{ij})$. 
This relationship enables us to identify an 
operator $\hat{\bf L}^{(ij)}(\hat{\omega}_{ij},\hat{k}_{ij})$ according to:
\begin{equation}
\hat{\bf L}^{(ij)}(\hat{\omega}_{ij},\hat{k}_{ij}) =
\left( 
i {\bf I} \sum_{i=1}^{N} \sum_{j=1}^{M} \hat{\omega}_{ij} +
\sum_{n=1}^{K} {\bf A}^{(n)} 
\left[ \sum_{i=1}^{N} \sum_{j=1}^{M} \hat{k}_{ij} 
\right]^{n} 
\right)
\label{L_opp2}
\end{equation}
If we moreover assume that the operator $Q$ is of the 
class $\hat{P}$ so that:
\begin{equation}
Q[{\bf u}(x,t)] =
\sum_{\mu_{1} \cdots \mu_{N}=1}^{\infty} 
\hat{{\bf R}}(\mu_{1} \cdots \mu_{N})
\prod_{i=1}^{N} \prod_{j=1}^{M}
f^{(i)}(x,t|\hat{k}_{ij},\hat{\omega}_{ij})
\label{Q_opp2}
\end{equation}
than the expansion coefficients are determined by the following linear
iteration series:
\begin{equation}
\hat{\bf L}^{(ij)}(\hat{\omega}_{ij},\hat{k}_{ij})
{\bf \hat{C}}(\mu_{1} \cdots \mu_{N})
=
{\bf \hat{R}}(\mu_{1} \cdots \mu_{N})
\label{lin_final}
\end{equation}
From this result we can conclude that we can transform Eq.(\ref{prob}) into
Eq.(\ref{lin_final}). We can conclude that an NPDE of the form Eq.(\ref{prob})
is integrable if the following two conditions are satisfied:
\begin{itemize}
\item{ The nonlinearity must have such a structure that it can be expanded 
       in the same basis functions as the linear part. In other 
       words, the nonlinearity must guarantee that Eq.(\ref{Q_opp2}) 
       is satisfied.} 
\item{ The inverse  the matrix 
       $\hat{\bf L}^{ij}(\hat{\omega}_{ij},\hat{k}_{ij})$ must exist}.  
\end{itemize}
From this result we can conclude that provided a solution (\ref{test1}) exits,
the integrability of the NPDE is completely determined by the linear 
part of the evolution equation.
These are also the conditions which guarantee 
the integrability of Eq.(\ref{prob}). In the following  section, 
we apply these concepts to examine the integrability of some NPDEs. 

\section{Examples}
In this section, we will apply the machinery developed in the precious 
sections to investigate the integrability of various NPDEs. As an first 
example, we consider the nonlinear Schr\"{o}dinger equation:
\begin{equation}
i \partial_{t} u = \partial_{xx} u+ 2 
u u^{\ast} u
\label{nls}
\end{equation}
If we substitute:
\begin{equation}
u(x,t) = e^{iax} e^{i(a^{2}-b^{2})t} e^{i \phi} 
\sum_{n=1}^{\infty} A_{n} e^{-n(bx-2abt)} 
\label{nls_sol}
\end{equation}
into Eq.(\ref{nls}), we obtain: 
\begin{equation}
\sum_{n=1}^{\infty}
\left[
(1 - n^{2}) b^{2}
\right]  
A_{n} e^{-n(bx-2abt)} 
= 
2 \
\sum_{n=1}^{\infty} \sum_{l=2}^{n-2} \sum_{m=1}^{n-l-1}
A_{l} A_{m} A_{n-m-l} e^{-n(bx-2abt)} 
\end{equation}
It can be verified that for $n=1$ the linear dispersion 
relationship $\omega = - k^{2} \ (k=a+bi)$ is satisfied. 
Since both the left-hand side and the right-hand-side can be 
expanded in the same Fourier basis-functions, we can determine 
the expansion coefficients by the following recursion relationship:
\begin{equation}
L^{(n)}(k) A_{n} = R^{(n)};
\hspace{1cm}
k = a + bi
\end{equation}
where
\begin{equation}
L^{(n)} = [1-n^{2}]b^{2};
\hspace{1cm}
R^{(n)} =
2 \sum_{l=2}^{n-2} \sum_{m=1}^{n-l-1}
A_{l} A_{m} A_{n-m-l};
\end{equation}
If we assume that $A_{1}=A$, than by computing all the coefficients $A_{n}$, 
and carrying out the summation, similarly as in Eq.(\ref{ser_soliton}), 
we obtain the NLS-soliton:
\begin{equation}
u(x,t) = 2 b e^{iax} e^{i(a^{2}-b^{2})t} e^{i \phi}
\mbox{sech}(bx - 2 abt + \xi_{0} )
\hspace{1cm}
\xi_{0} = \log \left( \frac{ | A | }{ 2 \eta } \right) 
\end{equation}
Similarly as for the KdV equation the two-soliton solution 
of the nonlinear Schr\"{o}dinger equation can be computed by considering 
solutions:
\begin{equation}
u(x,t) = 
e^{i ( a_{1} + a_{2} )  x } 
e^{i( a_{1}^{2} + a_{2}^{2} - b_{1}^{2} - b_{2}^{2} )t}
e^{i \phi}  
\sum_{n,m=1}^{\infty} 
C(n,m) e^{-n([ b_{1} + b_{2}] x- 2[ a_{1} b_{1} + a_{2} b_{2} ]t)}
\label{nls_sol1}
\end{equation}
By generalizing this procedure as presented in Sec.3, the N-soliton solution 
of the nonlinear Schr\"{o}dinger equation can be computed.

As a second example we consider the coupled nonlinear 
Schr\"{o}dinger equation:
\begin{equation}
\left\{
\begin{array}{l}
i u_{1t}  = u_{1xx} + (|u_{1}|^2 + |u_{2}|^2) u_{1} = 0 \\
i u_{2t}  = u_{2xx} + (|u_{2}|^2 + |u_{1}|^2) u_{2} = 0
\end{array} \right.
\label{coupnls}
\end{equation}
If we make the following substitution for the solution 
${\bf u}(x,t)= [ u_{1}(x,t),u_{2}(x,t) ]^{T}$:
\begin{equation}
{\bf u}(x,t) = e^{iax} e^{i(a^{2}-b^{2})t}  
\sum_{n=1}^{\infty} {\bf A}_{n} e^{-n(bx-2abt)} 
\hspace{1cm}
{\bf A}^{(n)} = ( A^{(n)}_{1},A^{(n)}_{2} )^{T}
\label{nls_sol2}
\end{equation}
into Eq.(\ref{coupnls}), it can be easily verified that both 
the left-hand side and the right-hand side of Eq.(\ref{coupnls}) can be 
expanded in the same basis-functions. This is due to the fact that
both $u_{1}(x,t)$ and $u_{2}(x,t)$ have the same dispersion relation
$\omega(k) = -k^{2}$. As a result, we can determine the expansion 
coefficients ${\bf A}^{(n)}$ by solving the following recursion 
relation
\begin{equation}
{\bf L}^{(n)}(k) {\bf A}^{(n)} = {\bf R}^{(n)}
\hspace{1cm}
k = a + bi
\end{equation}
where
\begin{equation}
{\bf L}^{(n)} = {\bf I} \left[ 1 - n^{2} 
\right] b^{2};
\hspace{.5cm}
{\bf R}^{(n)} = 
\sum_{l=2}^{n-2} \sum_{m=1}^{n-l-1}
\left(
\begin{array}{c}
A^{(l)}_{1} A^{(m)}_{1} A^{(n-m-l)}_{1} + A^{(l)}_{2} A^{(m)}_{2} 
A^{(n-m-l)}_{1} \\
A^{(l)}_{1} A^{(m)}_{1} A^{(n-m-l)}_{2} + A^{(l)}_{2} A^{(m)}_{2} 
A^{(n-m-l)}_{2}
\end{array}
\right);
\end{equation}
As a more general example, we consider the Manakov problem which plays a
role in optical fiber communication \cite{Berkhoer,Manakov}:
\begin{equation}
\left\{
\begin{array}{l}
i u_{1t} +  i \delta u_{1x} = u_{1xx} + (|u_{1}|^2 + |u_{2}|^2) u_{1} = 0 \\
i u_{2t} -  i \delta u_{2x} = u_{2xx} + (|u_{2}|^2 + |u_{1}|^2) u_{2} = 0
\end{array} \right.
\label{eq: manakov}
\end{equation} 
The Manakov equation (\ref{eq: manakov}) reduces to the nonlinear 
Schr\"{o}dinger equation if $\delta =0$. For a non-negative $\delta$
however, $u_{1}$ and $u_{2}$ have different linear dispersion relations. 
If we substitute:
\begin{equation}
{\bf u}(x,t) =  
e^{   i (a_{1}+a_{2}) x}
e^{ - i (a_{1}-a_{2}) \delta t} 
e^{   i (a^{2}_{1} + a^{2}_{2} - b^{2}_{1} - b^{2}_{2} ) t }
e^{ i \phi}  
\sum_{n,m=1}^{\infty} {\bf C}(n,m) 
e^{ - n b_{1} ( x - \delta t  + 2 a_{1} t )} 
e^{ - m b_{2} ( x + \delta t  + 2 a_{2} t)}
\label{mansol}
\end{equation}
into Eq.(\ref{nls}), we obtain that the linear
dispersion relationships are determined if the coefficients
$C_{1}(1,0)$ and $C_{2}(0,1)$ are unequal to zero. It can be verified
by substitution that in this case the following linear 
dispersion relations are valid:
\begin{equation}
\omega_{i}(k_{i}) = \delta k_{i} - k_{i}^{2}
\hspace{1cm}
k_{i} =  a_{i} + i b_{i} 
\hspace{1cm}
i \in \{1,2 \}
\end{equation}
In general the coefficients ${\bf C}(n,m)$ are determined by the following linear
algebraic relationship:
\begin{equation}
{\bf L}^{(n,m)}(k_{1},k_{2}) {\bf C}(n,m) = {\bf R}^{(n,m)}
\end{equation}
where
\begin{equation}
{\bf L}^{(n,m)}(k_{1},k_{2}) =
(1 - n^{2} ) b_{1}^{2} + (1 - m^{2} ) b_{2}^{2} 
+ 2i \left\{ 
m a_{1} b_{2} + n a_{2} b_{1} \mp m \beta_{2} \delta 
\right\} + 2 a_{1} a_{2} - 2 n m b_{1} b_{2}
\mp 2 \delta a_{2}
\end{equation}
and
\begin{equation}
\begin{split}
{\bf R}^{(n,n')}  = 
\sum_{l=2}^{n-2}   \sum_{l'=2}^{n'-2}
\sum_{m=1}^{n-l-1} \sum_{m'=1}^{n'-l'-1}
&\left[
\left(
\begin{array}{c}
C_{1}(l,l') C_{1}(m,m') C_{1}(n-m-l,n'-m'-l') \\
C_{1}(l,l') C_{1}(m,m') C_{2}(n-m-l,n'-m'-l')
\end{array} 
\right) 
\right. \\
&+
\left.
\left(
\begin{array}{c} 
C_{2}(l,l') C_{2}(m,m') C_{1}(n-m-l,n'-m'-l') \\
C_{2}(l,l') C_{2}(m,m') C_{2}(n-m-l,n'-m'-l')
\end{array}
\right)
\right]
\end{split}
\end{equation} 
In principle general solutions for the Manakov problem can be obtained by 
following the procedure as described in Section 3. It is interesting to 
conclude that although the Manakov problem has no soliton solutions, the 
existence of the solutions (\ref{mansol}) seem to suggest that the
the Manakov problem is integrable. The discussing about the integrability 
can be made general in a similar manner as presented in Section 2.  
The most simple solution of the Manakov
equation, however has four spectral parameters. This makes that this equation 
has no soliton solutions, but the most simple solution is an interacting
2-soliton solution. From a physical point of view, the Manakov problem describes 
the propagation of optical solitons over their principal birefringence 
axis. It is well know that optical solitons propagating through a birefringent medium
are unstable. This can be explained by the fact that initially localized optical
pulses disintegrate due to their soliton soliton interaction.  

As a last example, we consider the three dimensional nonlinear
Schr\"{o}dinger equation:
\begin{equation}
i \partial_{t} u = \sum_{n=1}^{3} \partial_{x_{n}}^{2} u+ 2 
u u^{\ast} u
\label{nls3}
\end{equation}
If we substitute:
\begin{equation}
u({\bf x},t) = e^{i{\bf a} \cdot {\bf x} } 
e^{i( {\bf a} \cdot {\bf a}- {\bf b} \cdot {\bf b})t} e^{i \phi} 
\sum_{n=1}^{\infty} A_{n} e^{-n({\bf b} \cdot {\bf x} -2 {\bf a} \cdot {\bf b}t)} 
\label{nls_sol3}
\end{equation}
into Eq.(\ref{nls}), we obtain: 
\begin{equation}
\sum_{n=1}^{\infty}
\left[
(1 - n^{2}) {\bf b} \cdot {\bf b}
\right]  
A_{n} 
e^{-n( {\bf b} \cdot {\bf x} -2 {\bf a} \cdot {\bf b}t)} 
= 
2 \
\sum_{n=1}^{\infty} \sum_{l=2}^{n-2} \sum_{m=1}^{n-l-1}
A_{l} A_{m} A_{n-m-l} 
e^{-n( {\bf b} \cdot {\bf x} -2 {\bf a} \cdot {\bf b}t)}
\label{sub3}
\end{equation}
In Eq.(\ref{nls_sol3}) and Eq.(\ref{sub3}), it is used that
${\bf x} = (x_{1},x_{2},x_{3})^{T}$, ${\bf a} = (a_{1},a_{2},a_{3})^{T}$
and ${\bf b} = (b_{1},b_{2},b_{3})^{T}$.
It can be verified that for $n=1$ the linear dispersion 
relationship $\omega^{2} = - {\bf k} \cdot {\bf k} 
({\bf k}={\bf a} +{\bf b}i)$ is satisfied. 
Since both the left-hand side and the right-hand-side can be 
expanded in the same Fourier basis-functions, we can determine 
the expansion coefficients by the following recursion relationship:
\begin{equation}
L^{(n)}({\bf k}) A_{n} = R^{(n)};
\hspace{1cm}
{\bf k} = {\bf a} + {\bf b}i
\end{equation}
where
\begin{equation}
L^{(n)}({\bf k}) = [1-n^{2}]{\bf b} \cdot {\bf b};
\hspace{1cm}
R^{(n)} =
2 \sum_{l=2}^{n-2} \sum_{m=1}^{n-l-1}
A_{l} A_{m} A_{n-m-l};
\end{equation}
Similarly as in the one-dimensional case, explicit solutions of the 
three dimensional nonlinear Schr\"{o}dinger equation can be obtained
by carrying out the summation of the expansion coefficients. The 
discussion can be made more general by using more general expansion 
functions, similarly as in Eq.(\ref{nls_sol1})

%

\section{Conclusions}
We have presented a novel method to test the integrability of large classes 
of NPDEs. We can conclude that the integrability of NPDEs depends on two 
important conditions. The first condition is that the NPDE must have 
a solution of the form (\ref{test1}), or more precisely, the nonlinearity
must have the property (\ref{Q_opp}). This ensures that the nonlinear
part of the NPDE can be expanded in the same basis functions as the 
linear part. If this is the case, the expansion coefficients of the
solution can be determined by a simple linear algebraic equation.

The machinery developed in this paper is derived by discussing 
the existence of N-soliton solutions of the KdV equation as an example. 
These results have lead to an insight which is used to present a 
general discussion with respect to the existence of N-soliton solutions
of the KdV equation.
In Section three the machinery is firstly generalized for NPDEs which 
have solutions that can be expanded in  Fourier basis-functions and later 
generalized to general basis-functions.
Secondly, the power of the method is demonstrated to derive solutions 
of the nonlinear Schr\"{o}dinger equation in a simple way. It is shown that we can
systematically derive large classes of solutions of the nonlinear 
Schr\"{o}dinger equation. We have also demonstrated that it is
possible to systematically derive solutions of the coupled
nonlinear Schr\"{o}dinger equation. Finally we derive solutions of 
the Manakov problem. The Manakov problem cannot be solved by 
inverse scattering techniques, however we have demonstrated that
large classes of solutions can be derived. 
It is interesting to remark that the most simple solution 
of the Manakov problem involves two independent complex spectral
parameters. This suggests that the most simple solution consists 
of two interacting solitons. The latter is related to the well
known instabilities  of the solutions of the Manakov problem. 
Finally it is indicated the the methods presented in this paper 
can be used to find solutions and to investigate the integrability
to higher-dimensional NPDEs.

As a concluding remark we want to state that the linearization principle 
as presented in this paper can be applied to large classes of NPDEs 
having a polynomial type of nonlinearity. It is well possible that this
concept can be generalized to much larger classes of nonlinear problems.
As an example we mention Ref.\cite{Dorren1}, in which it is shown 
that a similar approach can be used to solve inverse scattering problems.


\newpage
\begin{thebibliography}{99}

\bibitem{Zakharov}
V.E. Zakharov (Ed.), {\em What is integrability}, Springer-Verlag, 1990.

\bibitem{Eck1}
F. Calogero and W. Eckhaus, {\em Nonlinear evolution equations,
rescaling model PDEs and their integrability. I \& II}, Inverse
Problems {\bf 3}, 229-262 (1987); {\bf 4}, 11-33 (1988). 

\bibitem{Ablowitz}
M.J. Ablowitz and P.A. Clarkson, {\em Solitons, Nonlinear evolution
equations and inverse scattering}, Cambridge University Press,
Cambridge 1991. 

\bibitem{Dorren}
H.J.S. Dorren, {\em A linearizing transformation for the Korteweg-de Vries 
equationa; generalizations to higher-dimensional nonlinear partial
differential equations}, J. Math. Phys. {\bf 39}, 3711-3729 (1998).

\bibitem{Berkhoer}
A.L. Berkhoer and V.E. Zakharov, {\em Self exitation of waves with
different polarizations in nonlinear media}, 
Sov. Phys. JETP {\bf 31}, 486-490, 1970.

\bibitem{Manakov}
S.V. Manakov, {\em On the theory of two-dimensional stationary self-focusing 
of electromagnetic waves}, Sov. Phys. JETP {\bf 38}, 248-252, 1974.

\bibitem{Dorren1}
H.J.S. Dorren, {\em A novel approach to estimate the stability of the 
one-dimensional Marchenko equation}, preprint


\end{thebibliography}
\end{document}